\documentclass{Interspeech}

\interspeechcameraready

\title{Children's Voice Privacy: First Steps And Emerging Challenges}

\author[affiliation={1}]{Ajinkya}{Kulkarni}
\author[affiliation={2}]{Francisco}{Teixeira}
\author[affiliation={1}]{Enno}{Hermann}
\author[affiliation={2}]{Thomas}{Rolland}
\author[affiliation={2,3}]{Isabel}{Trancoso}
\author[affiliation={1}]{Mathew}{Magimai.-Doss}

\affiliation{Idiap Research Institute}{Martigny}{Switzerland}
\affiliation{INESC-ID, Lisbon, Portugal; $^{3}$Instituto Superior Técnico}{University of Lisbon}{Portugal}
\email{ajinkya.kulkarni@idiap.ch, francisco.s.teixeira@inesc-id.pt}

\usepackage[dvipsnames]{xcolor}

\usepackage[T1]{fontenc}
\usepackage{comment}
\usepackage{multirow}
\usepackage{booktabs}

\begin{document}

\maketitle

\begin{abstract}
Children are one of the most under-represented groups in speech technologies, as well as one of the most vulnerable in terms of privacy. Despite this, anonymization techniques targeting this population have received little attention. In this study, we seek to bridge this gap, and establish a baseline for the use of voice anonymization techniques designed for adult speech when applied to children's voices. Such an evaluation is essential, as children's speech presents a distinct set of challenges when compared to that of adults. This study comprises three children’s datasets, six anonymization methods, and objective and subjective utility metrics for evaluation. Our results show that existing systems for adults are still able to protect children's voice privacy, but suffer from much higher utility degradation. In addition, our subjective study displays the challenges of automatic evaluation methods for speech quality in children's speech, highlighting the need for further research.

\end{abstract}

\vspace{-0.1cm}
\section{Introduction}

The rapid evolution of technology over recent decades has significantly transformed children's activities, shifting many from physical spaces to digital environments. The increasing adoption of speech technologies underscores their importance in children's education, offering advanced learning tools such as interactive reading tutors and automated reading assessments that enhance language development and improve literacy through tailored speech processing \cite{c1a}. Recent studies\footnote{\url{https://www.ofcom.org.uk/media-use-and-attitudes/media-habits-children/}} indicate that nearly all children aged 3 to 17 in the UK engage with online content, with smartphones (72\%) and/or tablets (69\%) being the most commonly used devices. In the United States, approximately 12\% of voice assistant users are under the age of 12, highlighting that children represent a key demographic in the use of speech technology \cite{c24,kats}. Software developers often rely on third-party tools to create children's web applications, which frequently collect sensitive personal information. Given that children have a limited understanding of digital privacy risks compared to adults \cite{c52}, there is a growing need to implement robust measures that safeguard children's privacy rights and promote their autonomy in digital environments.

Several studies have consistently identified discrepancies in the performance of Automatic Speech Recognition (ASR) systems between children and adults due to challenges related to acoustic and language modeling \cite{f6,af8}. Children's speech processing presents unique challenges due to variations in pitch, pronunciation, vocal tract shape, and speech patterns, which differ from those of adults, making it difficult for speech recognition models to achieve high performance \cite{af8,a1}. The differences are primarily attributed to the physical and mental growth of children, reflecting diverse rates of speech development and changes in pronunciation as they mature. Additionally, children's speech often includes mispronunciations and disfluencies, further hindering accurate annotation and modeling~\cite{f14}. 

Although privacy regulations like COPPA and GDPR mandate technology companies to implement privacy policies \cite{gdpr}, limited research has been conducted in the context of children's speech privacy, particularly regarding the anonymization of children's speech ~\cite{dutta2025exploring,arasteh2024addressing,arasteh2024differential}. 
Early work by Bassiou et al.~\cite{bassiou2016privacy} focused on preventing the exposure of sensitive linguistic content by introducing a method to evaluate student collaboration using non-lexical speech features. Building on this concept, Dutta et al.~\cite{dutta2025exploring} recently explored the use of discrete speech units to enhance privacy and efficiency in speech recognition systems for pre-school and school-aged children. 
Arasteh et al.~\cite{arasteh2024addressing, arasteh2024differential} explored privacy challenges in pathological speech data, evaluating whether anonymized adult and children’s recordings remained valuable for the diagnosis of speech-affecting diseases, although not providing specific analyses for children alone.

This limited number of contributions shows that, while voice anonymization for adults is a growing area of research, in large part due to the Voice Privacy Challenges that have been organized since 2020~\cite{tomashenko2020voiceprivacy,tomashenko2024voice}, children's voice privacy, particularly anonymization, remains largely unexplored. 
In this work we aim to contribute to this under-explored research topic, and provide a set of baselines on how existing voice anonymization approaches can transfer to children's voices, and examine the challenges that arise in evaluating children's speech.
Specifically, the main contributions of this paper are:
\begin{itemize}
    \item Establishing a baseline of state-of-the-art voice anonymization techniques applied to children's speech data.
    \item Analyzing the performance of these techniques across multiple children's speech datasets and an adult speech dataset for comparison.
    \item Providing both objective and subjective evaluations of the voice anonymization approaches.
\end{itemize}
While our results show promise, they also reveal unique challenges inherent to children's voice anonymization, underscoring the need for tailored systems designed specifically to address the characteristics of children's speech.

\vspace{-0.1cm}
\section{Methodology}
\vspace{-0.1cm}

 




To explore the questions raised in the previous section, we examined different datasets combined with several voice anonymization systems, focusing on, as a first step, anonymizing children's speech to adult speech.
We analyzed the impact of anonymization across three children's speech datasets, each with distinct characteristics: MyST~\cite{ward2013my}, representing children's speech in educational scenarios; SpeechOcean762~\cite{so}, featuring non-native English-speaking children in language learning settings; and Samrómur~\cite{Samrmur}, an Icelandic children's speech dataset representing non-English speech.
We considered several baseline anonymization systems used in the Voice Privacy Challenge 2024 (VPC24), along with two voice conversion methods, and evaluated these methods using ASR and Automatic Speaker Verification (ASV) for utility and privacy, respectively.
Additionally, we used automatically predicted Mean Opinion Scores (MOS) and conducted a subjective listening study to ensure consistency between subjective and objective evaluations of voice anonymization systems.  
In the following sections we detail each of these experimental components.

\vspace{-0.2cm}
\subsection{Children's Speech Datasets}
\vspace{-0.1cm}

In this work, we used three children's voice datasets, namely My Science Tutor (MyST)~\cite{ward2013my}, Samromur \cite{Samrmur} and SpeechOcean762 \cite{so}, to provide a multi-domain investigation of children's voice privacy. In this section, we provide detailed descriptions of the datasets and the evaluation protocols designed to assess privacy aspects using ASV performance. In addition to these datasets, we also conducted baseline experiments using the LibriSpeech dataset~\cite{librispeech}, following the trial setup and anonymization systems provided by VPC24, to have an adult speech baseline.
\vspace{-0.3cm}
\paragraph*{MyST} The My Science Tutor (MyST) Children's Speech Corpus \cite{ward2013my} is one of the largest publicly available datasets of English children's speech, comprising approximately 400 hours of audio. The corpus captures conversations between children and a virtual tutor across eight scientific domains. Speech data were collected from 1,372 students in the third, fourth, and fifth grades, recorded at a sampling rate of 16 kHz. Since the age group of students in these class grades falls within a younger range, we opted against partitioning the MyST dataset for this study. Specifically, we selected 20 utterances per speaker, with a total of 50 speakers for both testing and imposter scenarios, taken from the test set of MyST, resulting in a dataset of 1,000 utterances and 
in 100k trials for ASV evaluation.

\vspace{-0.3cm}
\paragraph*{SpeechOcean762}
SpeechOcean762 \cite{so} is a dataset designed for pronunciation assessment, comprising 5,000 English utterances, sampled at 16 kHz, from 250 non-native Mandarin speakers, with half of the speakers being children. 
Each utterance was annotated at the sentence, word, and phoneme levels by five expert linguists.
This study specifically utilizes the test set of SpeechOcean762, comprising only children speakers. 
We selected 15 utterances per speaker for evaluation with 5 additional utterances used for enrollment. 
To ensure balanced representation, we selected two speakers from each age and divided them into two age groups for a more comprehensive analysis: Age Group 1: 6–10 years and Age Group 2: 11–15 years. This amounted to a total of 10 speakers and 150 test utterances,
For impostor utterances, we included both adult and child speakers. We further ensured that each speaker in the enrollment and impostor sets had a minimum of 5 utterances per speaker.
In total, we constructed 7050 trial pairs for ASV evaluation. 



\vspace{-0.3cm}
\paragraph*{Samrómur}
Samrómur Children dataset \cite{Samrmur} is an Icelandic speech corpus designed for ASR research. It contains 131 hours of read speech collected from 
children aged 4 to 17 years, sampled at 16 kHz. The dataset was compiled through crowdsourcing via samromur.is, a platform inspired by Mozilla’s Common Voice Project~\cite{Ardila2020}. 
%
For this study, we utilized a test set of Samrómur Children, excluding non-native Icelandic children's speech to ensure a more consistent linguistic dataset. To achieve balanced representation, we selected 2 speakers per age and categorized them into 2 distinct age groups: Age Group 1: 6–10 years and Age Group 2: 12–16 years. 
Each age group comprises 66 child speakers, with a minimum of 2 utterances per speaker. On average, approximately 13 speakers per age were selected, contributing 3 utterances per speaker, resulting in a total of 200 utterances. 
For impostor utterances, only child speakers were included. Additionally, we ensured that each speaker in the enrollment and impostor sets included at least 5 utterances. 
In total, we constructed 12,454 trial pairs for ASV evaluation.




\vspace{-0.2cm}
\subsection{Voice anonymization systems}
\vspace{-0.1cm}
In this work, we explored a diverse set of voice anonymization systems, to understand how the characteristics of each system affect its performance in the context of child voice anonymization. We included four of the baseline anonymization systems from the 2024 Voice Privacy Challenge (VPC24)~\cite{tomashenko2024voice}, in addition to two other voice conversion methods.
\vspace{-0.3cm}
\paragraph*{McAdams coefficient}
The first system uses McAdams coefficient anonymization (baseline B2 in VPC24), a signal processing-based approach that does not require pre-trained components. First proposed in \cite{patino2021speaker}, this approach aims to change the timbre of the speaker's voice, by modifying formant frequencies. Specifically, in this system: Linear Predictive Coding (LPC) source-filter analysis is applied to the speech signal, frame-wise; LPC coefficients are converted to poles in the z-plane; each non-real pole is raised to the power of the McAdams coefficient $\alpha$, to modify its phase and consequently the frequency of the formant; the modified poles are reconverted to LPC coefficients, which, in turn, are re-synthesised into speech. 

\vspace{-0.3cm}
\paragraph*{ASR-BN}
The second anonymization system corresponds to the VPC24 baseline system B5~\cite{tomashenko2024voice, champion2023anonymizing} -- an instance of an x-vector-based speaker anonymization system. 
In this system, the signal to be anonymized is decomposed into (1) F0, or pitch sequence and (2) vector-quantized bottleneck (BN) features extracted from a pre-trained ASR system. For anonymization, the signal is re-synthesized using a Hifi-GAN vocoder, which receives the bottleneck features, the F0 sequence, and a random one-hot encoding of the voice of one of the 247 adult speakers used in training the vocoder.

\vspace{-0.2cm}
\paragraph*{Phonetic transcriptions + GAN (STTTS)}
The third system is the approach of Meyer et al.~\cite{meyer2023prosody} (baseline B3 in VPC24). This system builds on x-vector based anonymization~\cite{srivastava2020design}, providing multiple improvements. The first improvement is the replacement of ASR bottleneck features with transcribed phonemes, effectively removing all speaker information contained in this branch. Secondly, pseudonymous x-vectors are generated through a Wasserstein - Generative Adversarial Network (GAN), to avoid using existing voices for anonymization. Thirdly, to keep the prosodic content, this system also includes a more complex prosodic branch, including phoneme durations as well as (randomly perturbed) pitch and energy. Finally, re-synthesis is performed using FastSpeech 2 and Hifi-GAN.

\vspace{-0.2cm}
\paragraph*{Neural Audio Codec (NAC)}
Our fourth system corresponds to the Neural Audio Codec (NAC) -- an encoder-decoder architecture, trained to represent audio signals as sequences of discrete tokens -- voice anonymization system of Panariello et al.~\cite{panariello2024speaker} (baseline B4 in VPC24). 
To perform voice anonymization, the authors proposed a two-branch system: in the first branch, a set of quantized semantic tokens that represent the linguistic content are extracted from the input signal; the second branch receives as input a pseudo-speaker prompt, chosen at random from the pool of training speakers, and generates NAC acoustic tokens, through the NAC encoder. Then, two transformers are used to combine the target acoustic tokens with the input semantic tokens, and to generate a new set of NAC acoustic tokens that reflect the semantic content of the input signal. The NAC decoder is then used to reconstruct the speech signal. 

\vspace{-0.3cm}
\paragraph*{KNN-VC and KNN-VC + Rhythm}
We further investigate two voice conversion methods based on \textit{k}-nearest neighbors (KNN). KNN-VC~\cite{Baas2023,Das2024} first extracts features for the source and target speaker audio from WavLM layer 6. Each source frame is then replaced with a frame of the target speaker that is selected with the KNN algorithm. Finally, a Hifi-GAN vocoder trained on WavLM features returns the output waveform. 
One limitation of KNN-VC, and other anonymization methods, is that it does not change the utterance length, so that speech rhythm and other prosodic aspects are not anonymized. As proposed in~\cite{hajal2025}, we combine KNN-VC with Urhythmic~\cite{urhythmic} (KNN-VC+R) to also modify the rhythm to that of a given target speaker. We use the pretrained WavLM and vocoder from~\cite{Baas2023}, while the conversion itself is non-parametric. Source speaker rhythm models are trained on their combined enroll and test utterances. For speakers with only a single utterance, we leave the rhythm unchanged due to insufficient data. 

\vspace{-0.3cm}
\paragraph*{Implementation details}
With the exception of the KNN-VC-based models, for all of the systems described above, we used the implementations provided by the organizers of the Voice Privacy Challenge 
using default configurations to perform utterance-level anonymization. All systems anonymize children's voices to adult voices, with the exception of the McAdams coefficient. KNN-VC and KNN-VC+R use as their target speaker the speaker from LJSpeech~\cite{ljspeech}.

\vspace{-0.2cm}
\subsection{Evaluation schema}
\vspace{-0.1cm}
We evaluate the performance of voice anonymization systems using objective and subjective measures. 
We use Equal Error Rate (EER) as our primary metric to measure privacy protection. EER scores are obtained using an ASV system. The evaluation of the ASV system assumes that the attacker has access to one trial utterance and several enrollment utterances. We used x-vectors computed with the ECAPA-TDNN system\footnote{\url{https://huggingface.co/speechbrain/spkrec-ecapa-voxceleb}} \cite{ecapatdnn} trained on the VoxCeleb 2 dataset~\cite{voxceleb2}. We obtained ASV scores by computing the cosine similarity between the mean speaker embedding of enrollment/impostor utterances and test utterances.
We assume a \textit{lazy-informed} attacker~\cite{srivastava2022privacy}, who has access to anonymized enrollment and trial utterances, but does not fine-tune the ASV model using anonymized data.

Our primary utility metric is the Word Error Rate (WER), obtained by performing ASR over the anonymized utterances, to assess the degradation of the linguistic content.
For English datasets, we perform ASR using Whisper large-v3 \footnote{\url{https://github.com/openai/whisper}}; for the Icelandic dataset, we use the Whisper-based model of Mena et al.~\cite{mena2024samromur}.
In addition to WER, we include WV-MOS scores \cite{wvmos} as a secondary utility metric for speech naturalness. The goal of this metric is to provide an indication of the degradation of the quality of the speech signal beyond the linguistic content.

Finally, we conduct a subjective listening test with the BeaqleJS framework~\cite{beaqlejs}. We randomly sampled five utterances from SpeechOcean and included these and the corresponding outputs from the six anonymization systems. We recruited 12~listeners who were asked to rate each of the 35 samples in terms of naturalness (\textit{whether the speech sounds like it’s produced by human} on a scale of 1 to~5) and to estimate the age of the speaker (0--10, 11--18, or older than 18~years).






\vspace{-0.1cm}
\section{Results and Discussion}
\vspace{-0.1cm}
\begin{table*}[ht]
\centering
\caption{Results in terms of EER and WER for all datasets. Org. refers to original speech; Higher EER $=$ better privacy, lower WER $=$ better intelligibility. The best results are indicated per dataset per metric in \textbf{bold}.}
\vspace{-0.2cm}
\resizebox{\textwidth}{!}{
\begin{tabular}{ll||c|cccccc||c|cccccc}
\toprule
\multicolumn{2}{c||}{\multirow{2}{*}{\textbf{Dataset}}} & \multicolumn{7}{c||}{\textbf{EER (\%)$\uparrow$}} & \multicolumn{7}{c}{\textbf{WER (\%)$\downarrow$}} \\ \cmidrule{3-16} 
                                 
& & \textbf{Org.} & \textbf{McAdams} & \textbf{ASR-BN} & \textbf{STTTS} & \textbf{NAC} & \textbf{KNN-VC} & \textbf{KNN-VC+R} 
& \textbf{Org.} & \textbf{McAdams} & \textbf{ASR-BN} & \textbf{STTTS} & \textbf{NAC} & \textbf{KNN-VC} & \textbf{KNN-VC+R} 
\\ \midrule
\multicolumn{2}{c||}{\textbf{LibriSpeech}}
& 0.41  & 25.24 & \textbf{49.49} & 46.94 & 46.05 & -- & --
& 2.56  & \textbf{3.89}  & 5.32  & 4.64  & 6.60  & -- & --
\\ \midrule
\multicolumn{2}{c||}{\textbf{MyST}}
& 5.0   & 23.40 & \textbf{47.80} & 46.50 & 44.40 & 28.49 & 30.27
& 13.42 & 25.32 & 27.40 & 59.23 & 24.44 & \textbf{22.33} & 24.61 \\ \midrule
\multirow{2}{*}{\textbf{SpeechOcean}} & AgeGrp1
& 4.67 & 18.0 & 55.33 & \textbf{52.0} & 55.33 & 30.67 & 35.33  
& 25.13 & 45.90 & 59.36 & 89.36 & 53.33 & \textbf{36.03} & 42.95 \\    
& AgeGrp2 
& 2.96 & 17.04 & 53.33 & \textbf{50.37} & \textbf{50.37} & 28.15 & 30.37  
& 15.05 & 39.25 & 38.82 & 37.31 & 41.72 & \textbf{30.54} & 39.25 \\ \midrule    
\multirow{2}{*}{\textbf{Samr\'omur}} & AgeGrp1
& 11.5 & 22.0 & 36.5 & 35.5 & \textbf{38.01} & 27.02 & 30.01  
& 11.48 & 40.96 & 104.78 & 144.92 & 86.69 & \textbf{40.24} & 45.83 \\    
& AgeGrp2 
& 5.10 & 20.41 & 30.61 & 32.65 & \textbf{33.16} & 17.86 & 18.88 
& 9.39 & 32.02 & 93.30 & 104.32 & 69.47 & \textbf{21.59} & 27.70 \\ \bottomrule 
\end{tabular}}
\label{tab:results_librimyst}
\vspace{-1.2em}
\end{table*}


\begin{table}[ht]
\centering
\caption{Results in terms of WV-MOS and ND-MOS for all datasets; Org. refers to original speech; higher MOS $=$ better naturalness.}
\vspace{-0.2cm}
\resizebox{\columnwidth}{!}{
\begin{tabular}{cll||c|cccccc}
\toprule
\multirow{2}{*}{\textbf{Metric}} & \multicolumn{2}{c||}{\multirow{2}{*}{\textbf{Dataset}}} & \multicolumn{7}{c}{\textbf{Anonymization System}}  \\ \cmidrule{4-10}
\multicolumn{3}{c||}{} &
\textbf{Org.} & \textbf{McAdams} & \textbf{ASR-BN} & \textbf{STTTS} & \textbf{NAC} & \textbf{KNN-VC} & \textbf{KNN-VC+R} \\ \midrule
\multirow{6}{*}{\textbf{WV-MOS}} & \multicolumn{2}{c||}{\textbf{LibriSpeech}} & 4.09  &  1.53 & 3.78  & \textbf{3.87}  & 3.71  & -- & -- \\ \cmidrule{2-10}
& \multicolumn{2}{c||}{\textbf{MyST}} & 2.79  & 0.60  & 3.11  & 2.97  & 2.83  & 3.49 & \textbf{3.64} \\ \cmidrule{2-10}
& \multirow{2}{*}{\textbf{SpeechOcean}} & AgeGrp1
& 3.03 & 1.33 & 3.41 & 3.70 & 3.20 & 3.69 & \textbf{3.88} \\
& & AgeGrp2 
& 3.28 & 1.51 & 3.58 & 3.99 & 3.51 & 3.88 & \textbf{4.19} \\ \cmidrule{2-10} 
& \multirow{2}{*}{\textbf{Samr\'omur}} & AgeGrp1
& 2.53 & 1.03 & 3.20 & 3.50 & 2.67 & 3.43 & \textbf{3.76} \\
& & AgeGrp2 
& 3.28 & 1.60 & 3.34 & 3.82 & 3.22 & 3.75 & \textbf{4.05} \\ \midrule\midrule
\textbf{ND-MOS} & \multicolumn{2}{c||}{\textbf{SpeechOcean}} 
& 3.87 & 2.27 & 2.42 & 2.24 & 2.66 & \textbf{3.27} & 3.15 \\
\bottomrule 

\end{tabular}}
\label{tab:results_librimyst_wvmos}
\vspace{-1.4em}
\end{table}


We start by discussing the results obtained for the MyST and LibriSpeech data, comparing the performance of the anonymization systems applied to children and adult speech\footnote{Speech samples of the  anonymization systems are available at \url{https://csp73896.github.io/}}.
These results can be found in the first two lines of Table \ref{tab:results_librimyst}, and show that, while existing anonymization systems are able to protect the privacy of children's speech to a level comparable to adults, as measured by EER, we can observe a stark degradation in terms of utility. Whereas for adult speech, WER suffers from an absolute degradation of at most 4\%, WER degradation for children's speech is on average $\sim$17\%, and over 45\% in the worst case (STTTS).
For McAdams, this degradation may be due to the use of hyperparameters that were optimized for adult speech, and it may be possible to conduct a hyperparameter search to improve utility. However, for the remaining systems this degradation is likely due to recognition errors in the linguistic branch of the anonymization systems, whose components are trained only on adult speech. Moreover, we hypothesize that the higher degradation of STTTS is due to the fact that this system utilizes phonetic transcriptions, corresponding to hard decisions, while the remaining neural-based systems use latent features.



Progressing our analysis to the two age-specific and out-of-domain datasets, SpeechOcean and Samr\'omur, we can observe similar trends for the results when comparing different systems to each other for both privacy and utility. Additionally, when considering individual age groups, the results show a trend towards lower WERs in age group two, which matches our expectations, as children in this group are older and thus have a speaking style that is closer to adults.
On the other hand, the privacy results for children in age group one are consistently higher. While this fact deserves further investigation, it may be possible that the poorer quality of the generated anonymized utterances may have a negative impact on ASV performance.

When analyzing the results by dataset, we can observe that privacy scores for Samr\'omur are generally lower when compared to MyST and SpeechOcean, whereas utility degradation is much higher. Both can potentially be justified by the presence of multiple language-dependent modules in the anonymization systems, which fail when presented with an unseen language.

Comparing different anonymization models, matching the literature~\cite{tomashenko2024voice}, McAdams provides the least privacy, having a large negative impact on utility. Moreover, while KNN-VC consistently obtains the lowest WER scores, it fails to provide a privacy level comparable to ASR-BN, STTTS or NAC. The best trade-offs between privacy and utility seem to be achieved by NAC, which consistently achieves high EER values, while presenting the least degradation, when compared to ASR-BN or STTTS. 
Although results from the literature show STTTS as providing a strong trade-off between privacy and utility~\cite{tomashenko2024voice}, the degradation introduced in terms of utility makes it a poor choice for children anonymization without prior adaptation.

In terms of WV-MOS, from Table \ref{tab:results_librimyst_wvmos}, we can observe that whereas for adult speech there is some slight degradation for most systems, for children's speech, with the exception of McAdams, and contrary to intuition, results improve. Our hypothesis is that the WV-MOS network is biased towards adult speech, as it did not encounter children's speech during training. Since all methods except McAdams anonymize children's voices to adult voices, the network may have a positive bias towards these, assigning them a higher quality than the original children's data. This hypothesis is further supported by the results obtained for SpeechOcean and Samr\'omur, wherein older age groups have higher WV-MOS scores. 

The final row of Table~\ref{tab:results_librimyst_wvmos} shows the naturalness MOS (ND-MOS) from the subjective listening test. The original speech samples are rated the highest, followed by KNN-VC, where rhythm modification only leads to a slight degradation. Listeners judged only 13\% of utterances anonymized with McAdams to sound like adult speakers, matching 12\% for the original samples as expected. For the remaining methods, which explicitly convert to adult speech, this value ranges from 73\% for NAC to 90--95\% for the others.

\vspace{-0.2cm}
\section{Challenges and Limitations}
\vspace{-0.1cm}

Despite the promising results in privacy protection, several challenges arise in applying voice anonymization techniques to children’s speech. One of the major challenges is identifying the exact sources of degradation in utility metrics. The decrease in ASR performance and speech quality may be attributed to multiple factors, including vocal tract differences, pitch variations, and dataset biases. Disentangling these confounding factors is necessary to develop more effective anonymization solutions. Another limitation lies in the potential misinterpretation of results. While EER improvements suggest enhanced privacy, the actual risk of re-identification in real-world scenarios remains uncertain. Particularly, in this work we only focus on \textit{lazy-informed} attackers, and it is possible that stronger attackers will result in reduced privacy guarantees. 
Conversely, WER degradation may overestimate the utility loss, as human listeners may still understand anonymized speech despite ASR errors. 

The obtained results also show the challenges in evaluating speech quality, specifically using WV-MOS, when compared with ND-MOS, this reflects the limitations in using techniques trained on adult speech and applying to children's speech. Existing datasets may also not fully capture the variability in children’s speech, leading to biased evaluations. On the other hand, transforming child speech into adult-like speech may limit the applications of anonymized voices, restricting usability in scenarios requiring child-specific speech characteristics. 

Additionally, a key limitation is that children's voices are extremely variable, and this study did not cover the full range of developmental stages or potential speech disorders present in the real world. Therefore, the outcomes of anonymization systems for a broader population of children remain unknown and should be explored as future work.



\vspace{-0.2cm}
\section{Conclusions and Future work}


This study establishes a preliminary understanding of children’s voice privacy by evaluating the performance of voice anonymization techniques originally developed for adult speech. Through experiments on diverse children’s speech datasets, the results indicate that anonymization to adult speech with existing systems can enhance privacy to certain degree. However, existing systems often degrade speech utility, particularly ASR performance and speech quality, due to factors such as non-native speech or linguistic variations (English \textit{vs} Icelandic). The age-specific analysis revealed that older children’s speech aligns more closely with adult-like speech patterns, leading to improved ASR performance, while younger children’s speech posed greater challenges for speech recognition. On the other hand, anonymization performance was found to be more dataset dependent.


Future research should address the identified challenges by improving anonymization techniques and tailoring them to children’s speech. Current anonymization systems primarily focus on adult voices. Adapting these systems to children requires optimizing ASR components, improving speaker embedding extraction, and refining voice conversion models. More robust pitch and prosody estimation methods tailored to children’s voices should be explored. 
In addition, following the research that is being conducted for adults, exploring anonymization methods that keep affect and prosodic content unchanged is also of importance, as it would greatly extend the number of applications of anonymized children's speech.

Instead of converting child voices to adult-like voices, a more suitable anonymization strategy may involve child-to-child voice transformations. However, this requires modifications in vocoder design, speaker embedding extraction, and ASR adaptation to ensure natural-sounding outputs while maintaining anonymity. 
Nevertheless, converting one child’s voice into another anonymized child-like voice may raise ethical concerns. Issues such as misrepresentation, identity falsification, and unintended biases in anonymized speech should be carefully examined. Ethical guidelines should be established to ensure that privacy-preserving techniques align with child protection policies. 
By addressing these challenges, future work can improve the effectiveness and applicability of voice anonymization systems for children, ensuring both privacy protection and high-quality speech usability.


\section{Acknowledgments}

This work was partially supported by the Swiss National
Science Foundation through the project PASS: Pathological Speech
Synthesis (grant agreement no. 219726) and by the Innosuisse through the flagship project IICT: Inclusive Information and Communication Technologies (grant agreement no. PFFS-21-47), and by the Portuguese national funds through Fundação para a Ciência e a Tecnologia, with reference DOI:10.54499/UIDB/50021/2020, by the Recovery and Resilience Plan and Next Generation EU European Funds, with reference C644865762-00000008 Accelerat.AI.
We thank André Mayoraz for setting up the listening test server.


\bibliographystyle{IEEEtran}
\bibliography{mybib}

\end{document}